\documentclass[aps,prb,twocolumn,secnumarabic,amsmath,amssymb,superscriptaddress]{revtex4-1}
\usepackage{dcolumn}
\usepackage{bm}

\usepackage{color} 
\usepackage{ulem}

\usepackage{amsmath}
\usepackage{graphicx}
\usepackage{float}
\usepackage{subfig}

\usepackage{color}
\usepackage[colorlinks,bookmarks=false,citecolor=blue,linkcolor=red,urlcolor=blue]{hyperref}

\definecolor{darkred}{rgb}{0.7,0.0,0.0}

\definecolor{darkblue}{rgb}{0,0.02,0.45}

\definecolor{darkgreen}{rgb}{0.02,0.45,0.0}

\definecolor{violet}{rgb}{0.8,0.2,0.6}

\providecommand{\U}[1]{\protect\rule{.1in}{.1in}}

\begin{document}

\title{Exact chiral spin liquid state in a Kitaev type spin model}

\author{Jianlong Fu}
\affiliation{School of Physics and Astronomy, University of Minnesota, Minneapolis, Minnesota 55455, USA}

\begin{abstract}
We study a frustrated two-dimensional lattice spin model with Kitaev type interaction. The lattice is obtained from the honeycomb lattice by replacing half of its sites with triangles. Using the SO(3) Majorana representation of spin, the model is exactly mapped into a $Z_{2}$ lattice gauge theory of complex fermions with standard Gauss law constraint. We show that the ground state of the model is a chiral spin liquid, and it has gapless excitations. The Chern number of the ground state is $\pm 1$, implying the existence of a chiral edge mode. The vortex excitation of the model is bound with a Majorana zero mode, such mode behaves as non-Abelian anyons.       
\end{abstract}

\maketitle
\section{Introduction}

Topological states of quantum many-body systems have attracted much interest in recent decades \cite{fradkinbook,schnyder08,chiu16,ludwig15,kitaev09,read00,kane051,kane052,qi11,hasan10,freedman04,tknn82,laughlin81,halperin82,Kitaev2003,Kitaev2006}. These states are usually characterized by an integer number (belonging to the group $Z$ or $Z_{2}$), states with different numbers cannot be smoothly deformed into each other. Topological states can support protected boundary modes \cite{fradkinbook,halperin82,kane051,kane052} and anyonic excitations \cite{Kitaev2006,wilczek1982,nayak2008}. For gapped free fermion systems and the general superconductor Bogoliubov-de Gennes (BdG) type of Hamiltonians, a complete classification of topological quantum states has been achieved \cite{schnyder08,chiu16,ludwig15,kitaev09}. But there are some difficulties in the classification of topological states with strong correlation \cite{chiu16}. Despite this, topological states exist in strongly correlated electronic systems. One distinct example is the chiral spin liquid states \cite{kalmeyer87,kalmeyer89,laughlin90,Wen1989,zou90}. In spin systems, chiral spin liquids are exotic ground states which break time reversal and parity symmetry. Such states were first proposed by Kalmeyer and Laughlin \cite{kalmeyer87,kalmeyer89} who noticed the similarity between certain frustrated spin Hamiltonians in hard-core boson representation and bosonic systems in certain magnetic field. They suggested that the ground states of these frustrated spin models can be described by the Laughlin wave function of bosonic quantum Hall states \cite{laughlin83}. Later, Wen, Wilczek and Zee proposed an order parameter for time-reversal and parity breaking spin states (chiral spin states) and constructed a Hamiltonian whose ground state is a chiral spin state \cite{Wen1989}. Ever since Kalmeyer and Laughlin, there has been a great effort in the study of chiral spin liquid states and trying to find Hamiltonians that support these states \cite{tsvelik17,wang17}. Based on the structure of the Laughlin wave function, parent Hamiltonians whose ground states are chiral spin liquids have been proposed \cite{schroeter07,thomale09,greiter09,greiter14}. These Hamiltonians generally involve complex spin interactions. Chiral spin liquid states have also been proposed in the kagome lattice systems \cite{yang93,chuaspin32,messio12,he14}.

Strongly correlated electronic systems are generally hard to study. For spin systems, exact results for models in two dimensions or higher are rare. One unique example of a two-dimensional exactly solvable spin model is the Kitaev honeycomb model \cite{Kitaev2006}. In this model, the bonds of the honeycomb lattice are characterized into $x,y$ and $z$ type; on a certain type of bond, the spin interaction is of Ising type with the corresponding spin component. Using Majorana representation of spin, the model can be mapped into a lattice $Z_{2}$ gauge theory in which fermions couple with $Z_{2}$ variables on bonds \cite{Kitaev2006,Baskaran2007,fu20181}. For each gauge sector, the Hamiltonian is a free Majorana fermion hopping model which can be transformed into a BdG Hamiltonian of complex fermions. The ground state of the Kitaev model is a $Z_{2}$ spin liquid \cite{Kitaev2006}. Quite remarkably, the exactly solvable Kitaev type of model can be introduced on other types of lattice, the only criteria is that there must be three bonds attached to each lattice site \cite{shankar09}. In particular, Yao and Kivelson proposed a Kitaev type spin model on the star lattice (also called the Fisher lattice) which comes from the honeycomb lattice by replacing every site with a triangle \cite{yao07}. It is shown that the ground state of this model is a chiral spin liquid, which agrees with the anticipation of Kitaev \cite{Kitaev2006}. By varying the relative strength of spin coupling on bonds, it is shown that there are two quantum phases in the model, called Abelian and non-Abelian phases \cite{yao07,chung10,kells10,shi10,nasu15,chua11,vidal08}. The Hamiltonian in this model is simple, which gives hopes of finding experimental realization. 

The ground state of the Kitaev type of spin model on star lattice \cite{yao07} is a topological quantum state, similar to certain phases of the Kitaev model itself \cite{Kitaev2006}. Here the difficulties in identifying topological states with strong correlation is overcome by the exact mapping between the spin model and the free fermion systems. Such mapping has been achieved by Majorana representation involving four Majorana fermions \cite{Kitaev2006} and one-dimensional Jordan-Wigner transformation \cite{yao07,feng07}. It is noteworthy that there is another spin representation that is suitable for the Kitaev type of models, the SO(3) Majorana representation \cite{Berezin1977,Tsvelik1992,fu20181,fu20182,Biswas2011,Shnirman2003,Mao2003,lai11}. Involving three Majorana fermions for each spin operator, the SO(3) Majorana representation can be applied to a class of spin models without SU(2) spin invariance, and exactly map these models into $Z_{2}$ lattice gauge theories with fermionic matter fields and standard Gauss law constraint \cite{fu20182}. Due to its non-local nature, it was also argued that the SO(3) Majorana representation is equivalent to the Jordan-Wigner transformation \cite{fradkinbook} in both one and two dimensions \cite{fu20182}. 

In order to further explore possible chiral spin liquid states that can appear in Kitaev type spin models as well as apply the SO(3) Majorana representation to spin states that break time-reversal symmetry, we propose and study the Kitaev type spin model on another type of lattice. The lattice is obtained from the honeycomb lattice by replacing half of its sites with triangles. We solve this model using the SO(3) Majorana representation of spin and exactly map the model into a lattice $Z_{2}$ gauge theory with standard Gauss law constraint. Starting from that, we show that its ground state is a chiral spin liquid, which agrees with Kitaev's anticipation that Kitaev type model defined on lattices which contain plaquettes with odd number of bonds can have ground states that spontaneously break time-reversal symmetry \cite{Kitaev2006}. We further explore the ground state of the model and show that it has Chern number $\pm 1$ which means that it has chiral edge modes. The spectrum of the bulk excitations is found to be gapless. We then explore the $Z_{2}$ vortex excitations of the model and show that there is a Majorana zero mode associated with each vortex, similar to the case of $p_{x}+ip_{y}$ topological superconductor \cite{read00,tewari07,chung07,roy10,ivanov01} and the Kitaev model itself \cite{Kitaev2006,lee07}. These Majorana zero modes behave as non-Abelian anyons \cite{nayak2008} when braiding among each other; therefore, the model may serve as a platform for quantum computation. Our model behaves differently from the similar model mentioned before proposed in Ref. \onlinecite{yao07} in that there is no quantum phase transition when varying the relative strength of spin coupling on different bonds.

The rest of the paper is organized as follows. In Sec. \ref{secmodelandsolution}, we introduce the Kitaev type spin model and its lattice. Using the SO(3) Majorana representation, we exactly map the original spin model into a $Z_{2}$ lattice gauge theory with fermionic matter fields and standard Gauss law constraints. In Sec. \ref{sectimereversalandspectrum}, we discuss the gauge structure of the $Z_{2}$ lattice gauge theory and analyze its properties under time reversal transformation. We find that the ground states of the model spontaneously break the time reversal symmetry, and the fermion spectrum under the assumption that the lattice translational symmetry is preserved is gapless. We conclude that the ground state of the model is a chiral spin liquid. In Sec. \ref{secchernandzeromode}, we study the topological properties of the model. We first compute the Chern number of the spectrum and find that it is equal to $\pm 1$. We then move on to discuss the flux excitation (or vortex) of the model and argue that each flux excitation is associated with a Majorana zero mode. Conclusion and some further discussion are given in Sec. \ref{secconclusion}.

\section{The model and its exact mapping to $Z_{2}$ lattice gauge theory} \label{secmodelandsolution}

\begin{figure}
\includegraphics[width=0.4\textwidth]{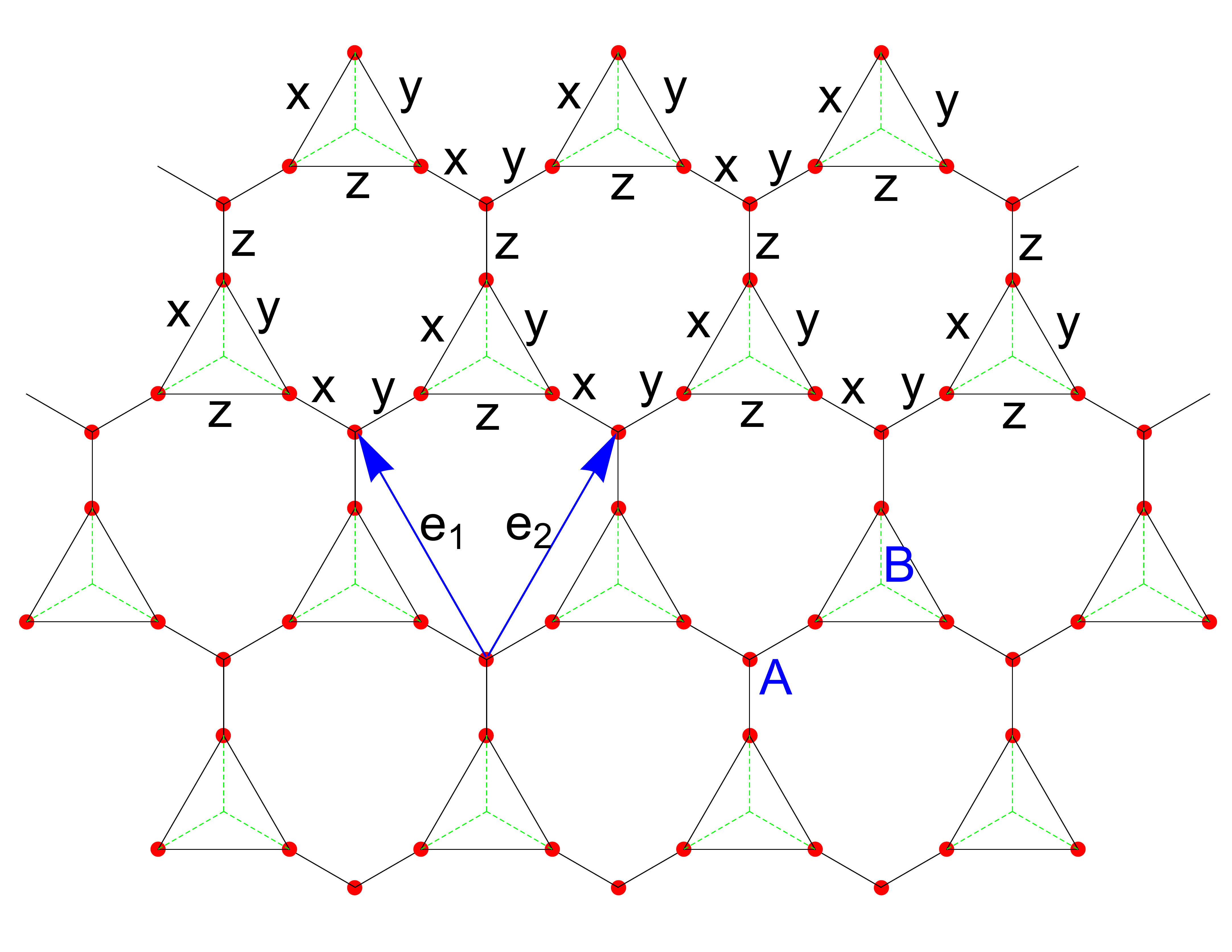}
\caption{The triangle-star lattice or wineglass lattice that the Kitaev type spin model is defined on. Each bond of the lattice is labeled by $x$, $y$ or $z$, only the corresponding spin component is interacting on each bond. The underlying honeycomb lattice has sublattices $A$ and $B$. The lattice vectors of the underlying honeycomb lattice are $\boldsymbol{e}_{1}$ and $\boldsymbol{e}_{2}$.}
\label{figlattice}
\end{figure}
   
\begin{figure}
\includegraphics[width=0.2\textwidth]{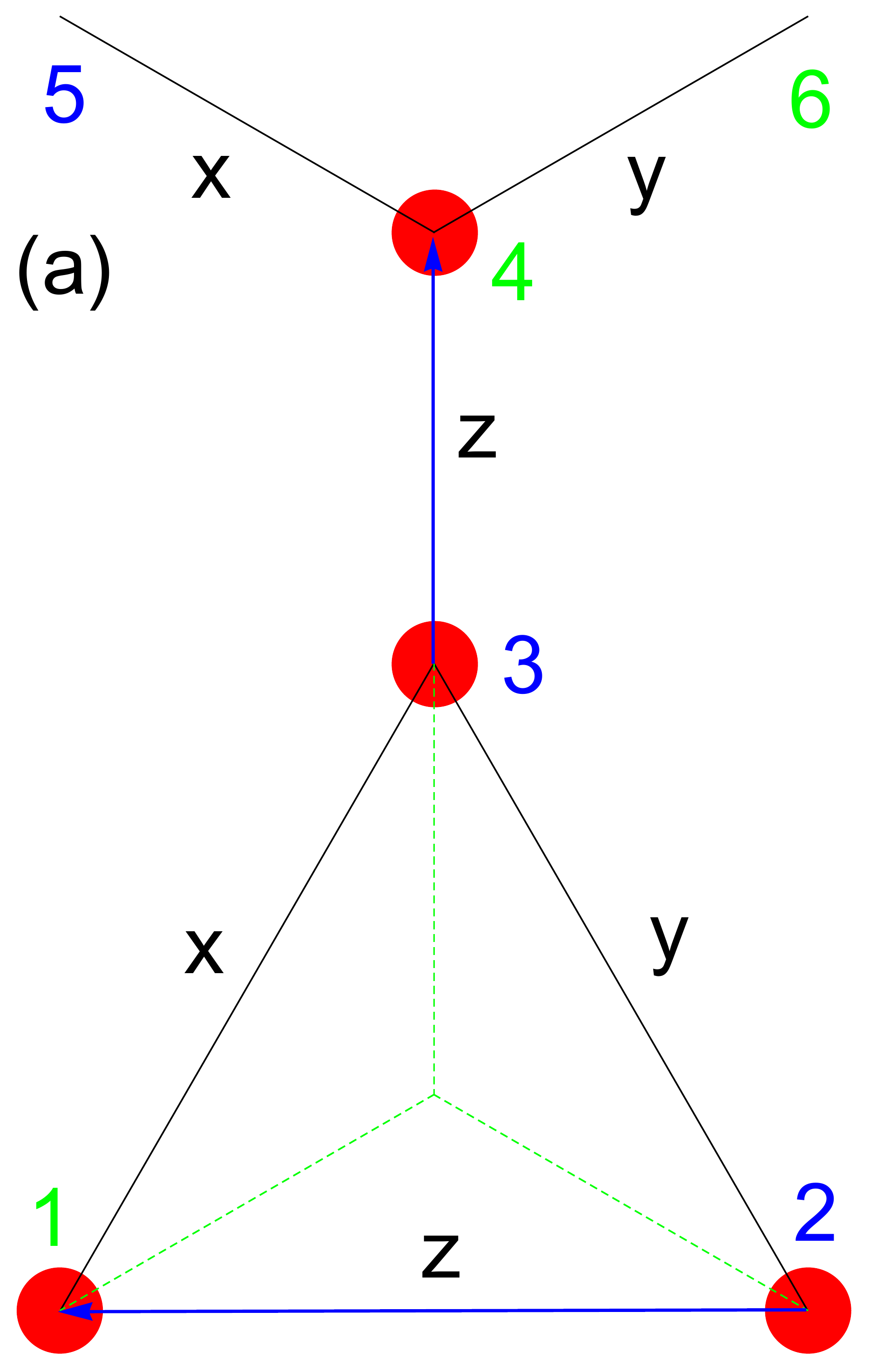}
\includegraphics[width=0.25\textwidth]{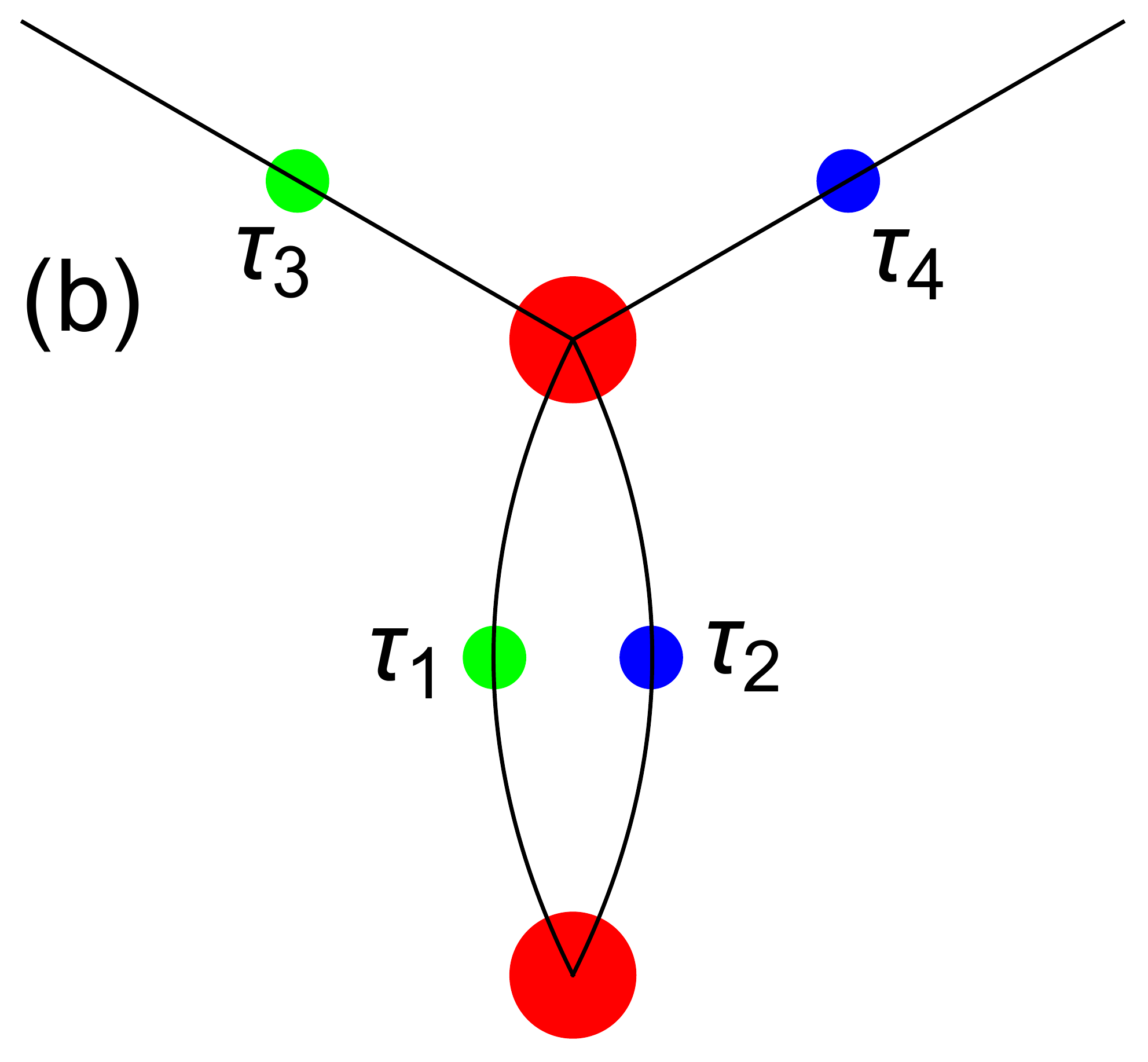}
\caption{The unit cell of (a) the original wineglass lattice, and (b) the effective lattice. In the unit cell of the original lattice (a), the sites are numbered 1 to 4, sites 5 and 6 belong to neighbouring unit cells. $z$-bonds $\langle 12\rangle$ and $\langle 34\rangle$ are paired with direction shown as blue arrows. In the unit cell of the effective lattice (b), there are two fermions and four $Z_{2}$ gauge fields $\tau_{1}$ to $\tau_{4}$. In both unit cells, the time reversal odd Majorana fermions (in (a)) and $Z_{2}$ gauge fields (in (b)) are labeled by blue, while the time reversal even objects are labeled by green.}
\label{figunit}
\end{figure}

To introduce the model, we consider a lattice which is obtained from the honeycomb lattice by replacing half of the sites with triangles, as shown in Fig. \ref{figlattice}. Each unit cell of the lattice consists of a triangle and a three-leg star (see Fig. \ref{figunit} (a)), the lattice can thus be called ``triangle-star" lattice or ``wineglass" lattice. Every bond of the lattice is labeled by $x$, $y$, and $z$ such that each type of bond appears once and only once around each site of the lattice (see Fig. \ref{figlattice}). The spin interaction is of the Kitaev type such that only the corresponding component of the spins on the two ends of every bond is interacting. The Hamiltonian of the model is given by
\begin{eqnarray}
\label{hamiltonian1}
\begin{aligned}
\mathcal{H}&=\sum_{\langle ij\rangle\in \triangle}J_{1}(\sum_{\langle ij\rangle_{x}}\sigma_{i}^{x}\sigma_{j}^{x}+\sum_{\langle ij\rangle_{y}}\sigma_{i}^{y}\sigma_{j}^{y}+\sum_{\langle ij\rangle_{z}}\sigma_{i}^{z}\sigma_{j}^{z})\\&+\sum_{\langle ij\rangle\in \bot}J_{2}(\sum_{\langle ij\rangle_{x}}\sigma_{i}^{x}\sigma_{j}^{x}+\sum_{\langle ij\rangle_{y}}\sigma_{i}^{y}\sigma_{j}^{y}+\sum_{\langle ij\rangle_{z}}\sigma_{i}^{z}\sigma_{j}^{z})\\
&=\sum_{\langle ij\rangle\in \triangle}\left(\sum_{\langle ij\rangle_{\alpha}}J_{1}\sigma_{i}^{\alpha}\sigma_{j}^{\alpha}\right)+\sum_{\langle ij\rangle\in \bot}\left(\sum_{\langle ij\rangle_{\alpha}}J_{2}\sigma_{i}^{\alpha}\sigma_{j}^{\alpha}\right),
\end{aligned}
\end{eqnarray} 
in which $\alpha=x,y,z$. Here and hereafter we use the symbols $\triangle$ and $\bot$ to denote the triangles and stars of the lattice respectively. The spin coupling constants of the triangle plaquettes and the stars are taken to be different and are denoted by $J_{1}$ and $J_{2}$ respectively. More generally, the coupling constants of each type of bond can be different from each other and the model is still exactly solvable, as with the Kitaev model \cite{Kitaev2006,fu20181}. Here for simplicity, we consider the case in which the coupling constants only depends on whether the bond is in triangle plaquettes or in stars.

To study this spin model we apply the SO(3) Majorana representation \cite{Berezin1977,Tsvelik1992,fu20181,fu20182,Biswas2011,Shnirman2003,Mao2003}. To this end, we introduce three Majorana fermions for each spin, these are denoted by $\eta^{x}$, $\eta^{y}$, and $\eta^{z}$. The SO(3) Majorana spin representation is defined by 
\begin{equation}
\label{so3first}
\sigma_{i}^{x}=-i\eta_{i}^{y}\eta_{i}^{z},\qquad \sigma_{i}^{y}=-i\eta_{i}^{z}\eta_{i}^{x},\qquad \sigma_{i}^{z}=-i\eta_{i}^{x}\eta_{i}^{y}.
\end{equation}
The three Majorana fermions on each site transform under the fundamental representation of SO(3), which corresponds to the SU(2) spin rotation. Using the three Majorana fermions, one can define a SO(3) singlet operator for each site 
\begin{equation}
\gamma_{i}=-i\eta_{i}^{x}\eta_{i}^{y}\eta_{i}^{z}.
\end{equation}
With the singlet the spin operator can also be written as
\begin{equation}
\label{so3second}
\sigma_{i}^{x}=\gamma_{i}\eta_{i}^{x},\qquad \sigma_{i}^{y}=\gamma_{i}\eta_{i}^{y},\qquad \sigma_{i}^{z}=\gamma_{i}\eta_{i}^{z}.
\end{equation}
Since the number of Majorana fermions on each site is odd, the Hilbert space of the Majorana fermions cannot be defined locally. In order to define the Majorana Hilbert space, one needs to pair up the sites \cite{fu20181,fu20182,Biswas2011}. Here we choose to pair up every $z$-bond of the lattice (in both the triangles and stars). For each paired $z$-bond, we enforce the following constraint to eliminate the extra dimensions of the Majorana Hilbert space \cite{fu20181,fu20182,Biswas2011} 
\begin{equation}
\label{constraint1}
\gamma_{i}\gamma_{j}=-i, \qquad \text{for all the $z$-bond $\langle ij\rangle_{z}$}.
\end{equation}
Here one has to specify the direction of the $z$-bond $\langle ij\rangle_{z}$. We choose that for each $z$-bond of the triangle, the direction is right-to-left and for each $z$-bond of the star, the direction is down-to-up; for both cases, the direction arrow is pointing to site $i$ of bond $\langle ij\rangle_{z}$ (see Fig. \ref{figunit} (a)). This specific choice of the bond direction does not influence the results, but it facilitates the definition. 
 
With these definitions, we can rewrite the spin Hamiltonian (\ref{hamiltonian1}) using the SO(3) Majorana representation. Here we use representation (\ref{so3first}) to represent spin interaction terms on the $x$-bonds and $y$-bonds, and use representation (\ref{so3second}) to represent spin interaction on the $z$-bonds, the spin Hamiltonian is thus transformed into
\begin{eqnarray}
\label{hamiltonian2}
\begin{aligned}
\mathcal{H}=\sum_{\langle ij\rangle_{\triangle}}\bigg[&\sum_{\langle ij\rangle_{x}}J_{1}(\eta_{i}^{y}\eta_{j}^{y})(\eta_{i}^{z}\eta_{j}^{z})+\sum_{\langle ij\rangle_{y}}J_{1}(\eta_{i}^{x}\eta_{j}^{x})(\eta_{i}^{z}\eta_{j}^{z})\\&+\sum_{\langle ij\rangle_{z}}J_{1}(-\gamma_{i}\gamma_{j})(\eta_{i}^{z}\eta_{j}^{z})\bigg]\\
+\sum_{\langle ij\rangle_{\bot}}\bigg[&\sum_{\langle ij\rangle_{x}}J_{2}(\eta_{i}^{y}\eta_{j}^{y})(\eta_{i}^{z}\eta_{j}^{z})+\sum_{\langle ij\rangle_{y}}J_{2}(\eta_{i}^{x}\eta_{j}^{x})(\eta_{i}^{z}\eta_{j}^{z})\\&+\sum_{\langle ij\rangle_{z}}J_{2}(-\gamma_{i}\gamma_{j})(\eta_{i}^{z}\eta_{j}^{z})\bigg].
\end{aligned}
\end{eqnarray}

For the next step, we pair up the $\eta^{z}$ Majorana fermions for the paired $z$-bonds and define complex fermion 
\begin{equation}
\label{complexfermiondefinition}
c_{i}=\frac{1}{2}(\eta_{i}^{z}-i\eta_{j}^{z}),\qquad c_{i}^{\dagger}=\frac{1}{2}(\eta_{i}^{z}+i\eta_{j}^{z}), 
\end{equation}
for each paired $z$-bond. Here the choice of $i$ and $j$ is made according to the direction of the $z$-bonds specified above. Conversely we have the Majorana fermion is written in terms of the complex fermions as
\begin{equation}
\eta_{i}^{z}=c_{i}+c_{i}^{\dagger}, \qquad \eta_{j}^{z}=i(c_{i}-c_{i}^{\dagger}).
\end{equation}
In these definitions we temporarily label the position of the complex fermions as the position of one of the Majorana fermions. 

\begin{figure}
\includegraphics[width=0.4\textwidth]{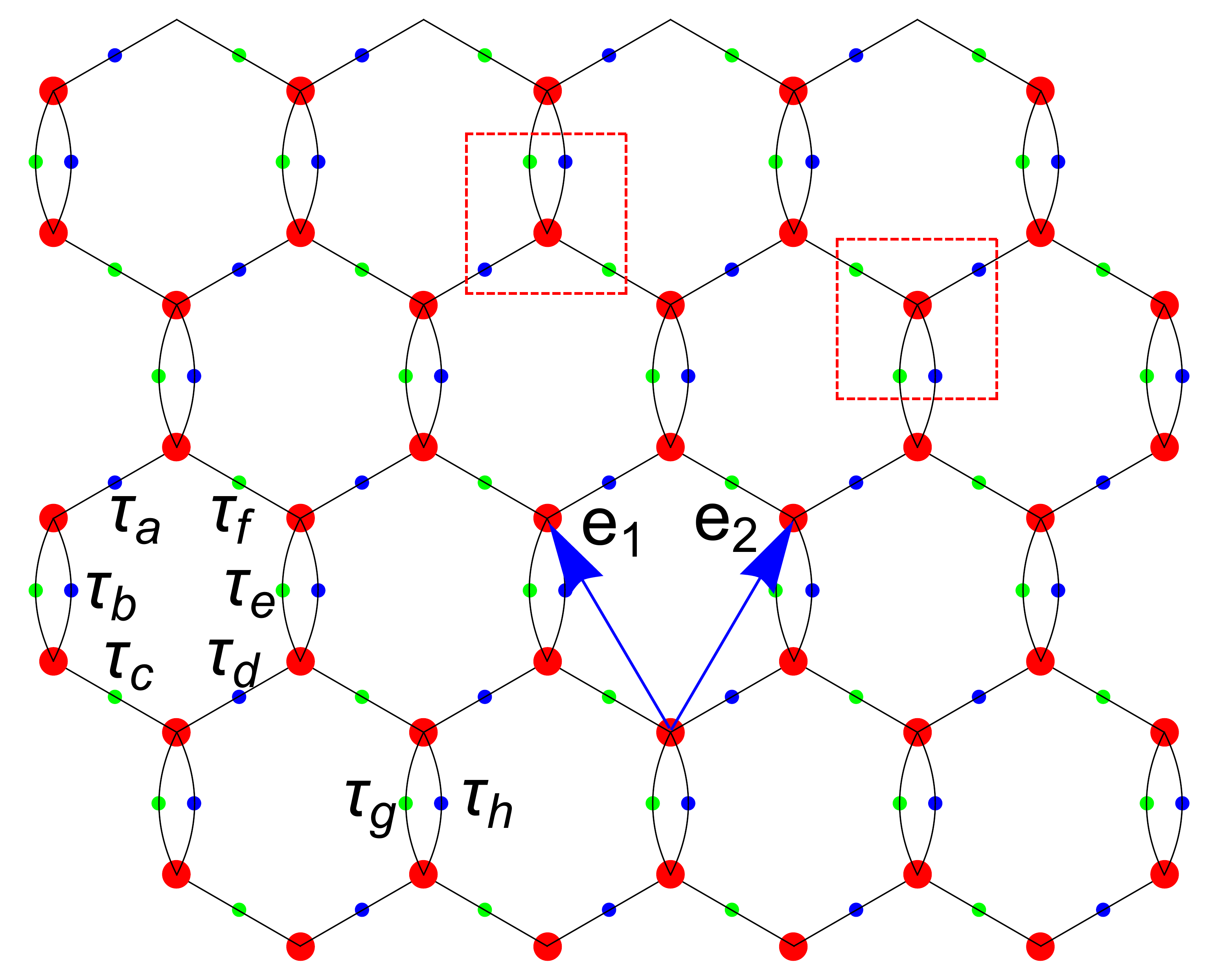}
\caption{The effective lattice of the model on which the $Z_{2}$ lattice gauge theory is defined. It has the shape of a honeycomb lattice. The fermions are located on the lattice sites (red dots), and the $Z_{2}$ gauge fields are labeled by the blue dots (time reversal odd) and green dots (time reversal even). The lattice vectors are given by $\boldsymbol{e}_{1}$ and $\boldsymbol{e}_{2}$. For each fermion site, the $Z_{2}$ Gauss law involves four adjacent gauge field operators (Eq. \ref{constraintz2condition}), these operators are enclosed in the red dashed boxes for two sublattices. The gauge fields $\tau_{a}$ to $\tau_{h}$ define two types of elementary plaquette terms in the model.}
\label{figeffective}
\end{figure}

After the pairing of all the $z$-bonds, the lattice has effectively become a honeycomb lattice (see Fig. \ref{figeffective}), with the triangles of the original lattice shrinked into their centers. The position of the complex fermion associated with each triangle $z$-bond is thus taken to be the center of the triangle. The $z$-bond of each star also shrinks to zero. The net result is that the complex fermions associated with the stars and the triangles are located on the A and B sublattices of the underlying honeycomb lattice respectively (see Fig. \ref{figeffective}). As shown by Fig. \ref{figeffective}, the $x$ and $y$-bonds of the triangle both effectively become the vertical bonds of the honeycomb lattice. Under this definition, we label the complex fermions associated with the stars as $c_{\hat{i}}^{A}$ and those associated with the triangle as $c_{\hat{i}}^{B}$, with $A$ and $B$ denoting the sublattices and $\hat{i}$ denoting the position of the unit cell in the effective honeycomb lattice.

To facilitate the discussion, we take the original unit cell, which consists of one triangle and one star. As shown in Fig. \ref{figunit} (a), we label the sites of the unit cell with numbers 1 to 6 (the sites of the triangle are 1,2,3 and the sites of the star are 3,4,5,6). Bonds $\langle 12\rangle$ and $\langle 34\rangle$ are $z$-bonds, with direction 2 to 1 and 3 to 4. Bonds $\langle 13\rangle$ and $\langle 45\rangle$ are $x$-bonds, Bonds $\langle 23\rangle$ and $\langle 46\rangle$ are $y$-bonds. With such labeling of sites in a unit cell and the definition of complex fermions (\ref{complexfermiondefinition}), we are able to rewrite the Hamiltonian (\ref{hamiltonian2}) as 
\begin{eqnarray}
\label{hamiltonian3}
\begin{aligned}
\mathcal{H}=\sum_{\hat{i}}\bigg[&iJ_{1}(\eta_{3}^{y}\eta_{1}^{y})(c_{\hat{i}}^{A}-c_{\hat{i}}^{A\dagger})(c_{\hat{i}}^{B}+c_{\hat{i}}^{B\dagger})\\&+(-J_{1})(\eta_{3}^{x}\eta_{2}^{x})(c_{\hat{i}}^{A}-c_{\hat{i}}^{A\dagger})(c_{\hat{i}}^{B}-c_{\hat{i}}^{B\dagger})\\&+(-J_{1})(2c_{\hat{i}}^{B\dagger}c_{\hat{i}}^{B}-1)\bigg]\\
+\bigg[&(-iJ_{2})(\eta_{5}^{y}\eta_{4}^{y})(c_{\hat{i}}^{A}+c_{\hat{i}}^{A\dagger})(c_{\hat{i}+\boldsymbol{e}_{1}}^{B}-c_{\hat{i}+\boldsymbol{e}_{1}}^{B\dagger})\\&+(-J_{2})(\eta_{6}^{x}\eta_{4}^{x})(c_{\hat{i}}^{A}+c_{\hat{i}}^{A\dagger})(c_{\hat{i}+\boldsymbol{e}_{2}}^{B}+c_{\hat{i}+\boldsymbol{e}_{2}}^{B\dagger})\\&+(-J_{2})(2c_{\hat{i}}^{A\dagger}c_{\hat{i}}^{A}-1)\bigg],
\end{aligned}
\end{eqnarray}
in which the site $\hat{i}$ labels the unit cell of the effective honeycomb lattice, and we have used the constraint (\ref{constraint1}). Specifically, the constraints that for every $z$-bond, Eq. (\ref{constraint1}) is satisfied, means that for every unit cell of the original lattice \cite{fu20181,fu20182},
\begin{equation}
\label{constraint2}
(-1)^{n_{\hat{i}}^{B}}(\eta_{1}^{x}\eta_{2}^{x})(\eta_{1}^{y}\eta_{2}^{y})=1, \quad (-1)^{n_{\hat{i}}^{A}}(\eta_{4}^{x}\eta_{3}^{x})(\eta_{4}^{y}\eta_{3}^{y})=1,
\end{equation}
in which $n_{\hat{i}}^{\alpha}=c_{\hat{i}}^{\alpha\dagger}c_{\hat{i}}^{\alpha}$ (with $\alpha=\text{A or B}$) denotes the number of complex fermion in unit cell $\hat{i}$ (of the effective honeycomb lattice) of sublattice A or B. These constraints commute with the Hamiltonian due to the properties of the SO(3) singlet operators in their original definitions (\ref{constraint1})\cite{fu20181,fu20182}.

For the next step we notice that the Majorana bilinear bond operators that appear in the Hamiltonian (\ref{hamiltonian3}) for each unit cell commute with each other and with those in the neighbouring unit cells. Furthermore, all the bond operators commute with the Hamiltonian. These allow us to intepret these bond operators as independent $Z_{2}$ variables in the Hamiltonian. Specifically we define the following auxiliary spin variables for bond operators in every unit cell,
\begin{eqnarray}
\begin{aligned}
\label{z2variables}
&\tau_{1}^{z}=-i\eta_{3}^{y}\eta_{1}^{y}, \qquad \tau_{2}^{z}=-i\eta_{3}^{x}\eta_{2}^{x},\\  &\tau_{3}^{z}=-i\eta_{5}^{y}\eta_{4}^{y},\qquad \tau_{4}^{z}=-i\eta_{6}^{x}\eta_{4}^{x}.
\end{aligned}
\end{eqnarray}
Note that in the definition above, on the right-hand-side of the equations we use the numbering of the unit cell of the original wineglass lattice; on the left-hand-side, the numbering of the $Z_{2}$ variables are given in the unit cell of the effective honeycomb lattice (see Fig. \ref{figunit} (b)). With these auxiliary bond $Z_{2}$ variables, the Hamiltonian (\ref{hamiltonian3}) is transformed into
\begin{eqnarray}
\label{hamiltonian4}
\begin{aligned}
\mathcal{H}=\sum_{\hat{i}}\bigg[&-J_{1}\tau_{1}^{z}(c_{\hat{i}}^{A}-c_{\hat{i}}^{A\dagger})(c_{\hat{i}}^{B}+c_{\hat{i}}^{B\dagger})\\&+(-iJ_{1})\tau_{2}^{z}(c_{\hat{i}}^{A}-c_{\hat{i}}^{A\dagger})(c_{\hat{i}}^{B}-c_{\hat{i}}^{B\dagger})\\&+(-J_{1})(2c_{\hat{i}}^{B\dagger}c_{\hat{i}}^{B}-1)\bigg]\\
+\bigg[&J_{2}\tau_{3}^{z}(c_{\hat{i}}^{A}+c_{\hat{i}}^{A\dagger})(c_{\hat{i}+\boldsymbol{e}_{1}}^{B}-c_{\hat{i}+\boldsymbol{e}_{1}}^{B\dagger})\\&+(-iJ_{2})\tau_{4}^{z}(c_{\hat{i}}^{A}+c_{\hat{i}}^{A\dagger})(c_{\hat{i}+\boldsymbol{e}_{2}}^{B}+c_{\hat{i}+\boldsymbol{e}_{2}}^{B\dagger})\\&+(-J_{2})(2c_{\hat{i}}^{A\dagger}c_{\hat{i}}^{A}-1)\bigg].
\end{aligned}
\end{eqnarray}
The $Z_{2}$ varibles connecting two complex fermion sites are placed on the bonds of the effective lattice, as shown by Fig. \ref{figeffective}; they are thus also referred to as ``bond spins". The bond spins are not free parameters, they do not commute with the constraints (\ref{constraint2}). To understand it better, we need to map the constraints (\ref{constraint2}) into a form with complex fermion operators and the $Z_{2}$ variables.

In order to further understand the role of the constraints, we first try to find a way to define the conjugate varible $\tau^{x}$ using the Majorana fermions. Take the $\tau^{z}$ on $x$-bonds for example, they are defined as a bilinear of $\eta^{y}$ Majorana fermions in (\ref{z2variables}). The definition of its corresponding $\tau^{x}$ operator can be obtained from a product of $\eta^{y}$ Majorana fermion on a chain of $z$-$x$-bonds. The chain can have finite length if a reference point is picked up along itself. We will not discuss the procedure in detail for this model for notation simplicity, but we will discuss it for the Kitaev model \cite{Kitaev2006,fu20181} in the Appendix \ref{appendixkitaev}. The process illustrated there can be directly applied to this model. After the definition of $\tau^{x}$ operators using a chain of Majorana fermions, we find that the constraints (\ref{constraint2}) can be written as
\begin{equation}
\label{constraintz2condition}
(-1)^{n_{\boldsymbol{r}}}\prod_{\boldsymbol{r}'\in \partial \boldsymbol{r}}\tau^{x}_{\boldsymbol{r}'}=1.
\end{equation}
In Eq. \ref{constraintz2condition}, we use the position vector $\boldsymbol{r}$ to label the position of the complex fermion in the effective honeycomb lattice, the site $\boldsymbol{r}$ can be either A or B sublattice site. Vector $\boldsymbol{r}'$ is used to label the position of the effective $Z_{2}$ varibles on bonds of the effective lattice. The product of $\tau^{x}$ operators are taken over the four $Z_{2}$ varibles surrounding the corresponding fermion sites. In Fig. \ref{figeffective}, the four $Z_{2}$ variables involved for fermions on two sublattices are enclosed in the red dashed boxes. 

The Hamiltonian (\ref{hamiltonian4}) with constraints (\ref{constraintz2condition}) defines a lattice $Z_{2}$ gauge theory \cite{Kogut1979,fradkinbook}. The constraints (\ref{constraintz2condition}) take the form of standard Gauss law \cite{fradkinbook}. These results agree with the prediction of Ref. \onlinecite{fu20182} that the application of the SO(3) Majorana representation usually leads to $Z_{2}$ lattice gauge theories. It is important to note that the mapping from the original spin Hamiltonian (\ref{hamiltonian1}) to the $Z_{2}$ lattice gauge theory is exact. It is noteworthy that under specific definitions of $\tau^{x}$ operators, the right-hand-side of Eq. \ref{constraintz2condition} can be $-1$ for some fermions. But this has no influence on the results provided that we change all the relevant definitions accordingly.

\section{Time reversal symmetry breaking and the complex fermion spectrum} \label{sectimereversalandspectrum}

\subsection{Gauge structure and time reversal symmetry breaking}

Before trying to find the eigenstates of the $Z_{2}$ gauge theory, we have to discuss its gauge structure. Eigenstates of the Hamiltonian (\ref{hamiltonian4}) itself can be written as $|\{\tau^{z}\}\rangle\otimes|\Psi_{E}\rangle_{\{\tau^{z}\}}$, in which $\{\tau^{z}\}$ denotes a distribution of the $Z_{2}$ gauge fields throughout the lattice and $|\Psi_{E}\rangle_{\{\tau^{z}\}}$ denotes the fermion state associated with $\{\tau^{z}\}$ determined by the Hamiltonian (\ref{hamiltonian4}) with energy $E$. These states form a complete basis of the eigenstates of the Hamiltonian (\ref{hamiltonian4}), therefore we will refer to these states as ``basis states". The Gauss law constraint means that the physical states are gauge invariant states. In other words, physical states are superpositions of all the basis states belonging to the same gauge sector.
The Gauss law (\ref{constraintz2condition}) defines the $Z_{2}$ gauge transfomation operator around fermion site $\boldsymbol{r}$ on its left-hand-side, namely $\mathcal{D}_{\boldsymbol{r}}=(-1)^{n_{\boldsymbol{r}}}\prod_{\boldsymbol{r}'\in \partial \boldsymbol{r}}\tau^{x}_{\boldsymbol{r}'}$. From these operators one can define a projector operator $\hat{\mathcal{P}}$,
\begin{equation}
\hat{\mathcal{P}}=\prod_{\boldsymbol{r}}\left(\frac{1+\mathcal{D}_{\boldsymbol{r}}}{2}\right).
\end{equation} 
By gauge invariance, the projector $\hat{\mathcal{P}}$ commutes with the Hamiltonian (\ref{hamiltonian4}). Using the projector operator, physical eigenstates with energy $E$ associated with basis state $|\{\tau^{z}\}\rangle\otimes|\Psi_{E}\rangle_{\{\tau^{z}\}}$ in the model can be written as
\begin{equation}
|\psi_{E}\rangle_{\text{phys}}\sim \hat{\mathcal{P}}\left[|\{\tau^{z}\}\rangle\otimes|\Psi_{E}\rangle_{\{\tau^{z}\}}\right],
\end{equation}
up to some normalization coefficient. These physical states correspond to the eigenstates of the original spin model. On the other hand, some general remarks can be made about the gauge structure based on the geometry of the effective lattice (see Fig. \ref{figeffective}). Most importantly, the $\tau_{1}^{z}$ and $\tau_{2}^{z}$ fields in any unit cell of the effective lattice can only change sign together under gauge transformations. This implies that $\tau_{1}^{z}\tau_{2}^{z}$ is a gauge invariant object in a certain unit cell (see Fig. \ref{figunit}). In other words, basis states with $\tau_{1}^{z}\tau_{2}^{z}=1$ of any unit cell belongs to different gauge sectors from states with $\tau_{1}^{z}\tau_{2}^{z}=-1$ of the same unit cell.

To study the time reversal symmetry in this model, we note that as a matter field, the complex fermion $c_{\hat{i}}^{\alpha}$ (or $c_{\boldsymbol{r}}$) should not change under time reversal transformation. Therefore, the Majorana fermions have to be either even or odd under time reversal (in other words, purely real or purely imaginary). From now on, we use $\hat{\mathcal{T}}$ to denote the antiunitary time reversal transformation. The original spin operator in (\ref{so3first}) are odd under $\hat{\mathcal{T}}$, which implies that the three Majorana fermions on a given site are either all even or all odd under $\hat{\mathcal{T}}$. Based on our specific choice of pairing on each $z$-bond, we have that in a single unit cell (see Fig. \ref{figunit}), the Majorana fermions transform in the following way under time reversal,
\begin{eqnarray}
\begin{aligned}
&\eta_{3}^{\alpha}\rightarrow -\eta_{3}^{\alpha}, \qquad \eta_{2}^{\alpha}\rightarrow -\eta_{2}^{\alpha}, \qquad \eta_{5}^{\alpha}\rightarrow -\eta_{5}^{\alpha}. \\
&\eta_{4}^{\alpha}\rightarrow \eta_{4}^{\alpha}, \qquad \eta_{6}^{\alpha}\rightarrow \eta_{6}^{\alpha}, \qquad \eta_{1}^{\alpha}\rightarrow \eta_{1}^{\alpha}. 
\end{aligned}
\end{eqnarray}
For the $Z_{2}$ gauge fields we have 
\begin{eqnarray}
\label{timereversalofz2}
\begin{aligned}
&\hat{\mathcal{T}}\tau_{1}^{z}\hat{\mathcal{T}}^{-1}=\tau_{1}^{z}, \qquad \hat{\mathcal{T}}\tau_{3}^{z}\hat{\mathcal{T}}^{-1}=\tau_{3}^{z}, \\ &\hat{\mathcal{T}}\tau_{2}^{z}\hat{\mathcal{T}}^{-1}=-\tau_{2}^{z}, \qquad \hat{\mathcal{T}}\tau_{4}^{z}\hat{\mathcal{T}}^{-1}=-\tau_{4}^{z}.
\end{aligned}
\end{eqnarray}
This gives the transformation of all the fields in the $Z_{2}$ gauge theory. Note that the Gauss law condition (\ref{constraintz2condition}) is invariant under time reversal and thus the projector operator $\hat{\mathcal{P}}$ commutes with time reversal $\hat{\mathcal{T}}$.

Among the $Z_{2}$ gauge fields in a single unit cell, $\tau_{2}^{z}$ and $\tau_{4}^{z}$ are time reversal odd; $\tau_{1}^{z}$ and $\tau_{3}^{z}$ are time reversal even. Consequently, the gauge invariant object $\tau_{1}^{z}\tau_{2}^{z}$ in a unit cell is odd under time reversal. Using this we conclude that for every basis states $|\{\tau^{z}\}\rangle\otimes|\Psi_{E}\rangle_{\{\tau^{z}\}}$, its time reversal partner $\hat{\mathcal{T}}|\{\tau^{z}\}\rangle\otimes|\Psi_{E}\rangle_{\{\tau^{z}\}}$ belongs to a different gauge sector. Therefore the physical eigenstate $|\psi_{E}\rangle_{\text{phys}}=\hat{\mathcal{P}}|\{\tau^{z}\}\rangle\otimes|\Psi_{E}\rangle_{\{\tau^{z}\}}$ and its time reversal partner $\hat{\mathcal{T}}|\psi_{E}\rangle_{\text{phys}}=\hat{\mathcal{P}}\hat{\mathcal{T}}|\{\tau^{z}\}\rangle\otimes|\Psi_{E}\rangle_{\{\tau^{z}\}}$ are different states. Because the basis states form a complete basis of all the eigenstates of the Hamiltonian (\ref{hamiltonian4}), the conclusion that $\hat{\mathcal{T}}|\psi_{E}\rangle_{\text{phys}}\neq |\psi_{E}\rangle_{\text{phys}}$ is also true for the state with the lowest eigenvalue $E_{g}$: $|\psi_{E_{g}}\rangle_{\text{phys}}$, which is a ground state of the Hamiltonian. Furthermore, the Hamiltonian (\ref{hamiltonian4}) is invariant under time reversal, which means that $\hat{\mathcal{T}}|\psi_{E_{g}}\rangle_{\text{phys}}$ is another eigenstates with the same energy. Thus we reach the conclusion that the ground state of the model spontaneously breaks time reversal symmetry.

The key element for arguing the spontaneously broken time reversal symmetry in $Z_{2}$ gauge theories is that the time reversal partner of each basis state belongs to a different gauge sector. Other types of $Z_{2}$ lattice gauge theories from the application of the SO(3) Majorana representation may not have such properties. Take the Kitaev model as an example, similar procedures leads to lattice $Z_{2}$ gauge theory with Gauss law constraint \cite{fu20181}. The $Z_{2}$ gauge fields in the Kitaev model are all even under time reversal, thus the ground state of the Kitaev model is invariant under time reversal.

The Hamiltonian (\ref{hamiltonian4}) is not invariant under interchange between $\boldsymbol{e}_{1}$ and $\boldsymbol{e}_{2}$, thus the ground state of the model also breaks parity symmetry. For the same reason as the Kitaev model, the spin-spin correlation functions vanish beyond nearest neighbouring sites \cite{Baskaran2007}. Therefore we conclude that the ground state of the model is a chiral spin liquid.

\subsection{Complex fermion spectrum}

In order to solve for the complex fermion spectrum, we only consider the states that are gauge equivalent to the states in which the distribution of the $Z_{2}$ gauge fields $\{\tau^{z}\}$ has the lattice translational symmetry of the effective honeycomb lattice. These states form a subset of the complete basis states $|\{\tau^{z}\}\rangle\otimes|\Psi_{E}\rangle_{\{\tau^{z}\}}$. Furthermore, we make an assumption that the ground state of the model is in this subset. A discussion on the gauge structure of these states is in order. We note that under the assumption that the states have lattice translational symmetry, the configurations of the $Z_{2}$ gauge fields throughout the lattice depend only on the $\tau_{1}^{z}$, $\tau_{2}^{z}$, $\tau_{3}^{z}$, $\tau_{4}^{z}$ in one unit cell. There are sixteen possibilities in total. It can be shown that under suitable gauge transformations, all the possible combinations of $\tau_{3}^{z}$ and $\tau_{4}^{z}$ are equivalent, with the same $\tau_{1}^{z}$ and $\tau_{2}^{z}$. However, those states with $\tau_{1}^{z}\tau_{2}^{z}=1$ and those with $\tau_{1}^{z}\tau_{2}^{z}=-1$ are not gauge equivalent. So the sixteen possibilities of $\tau_{1}^{z}$ to $\tau_{4}^{z}$ fall into two gauge sectors with $\tau_{1}^{z}\tau_{2}^{z}=\pm 1$. These two gauge sectors transform into each other under time reversal.

Under these considerations, we can work out lattice translational symmetric solutions to the Hamiltonian. We first perform Fourier transformation to the complex fermion fields,
\begin{equation}
c_{\hat{i}}^{\alpha}=\frac{1}{\sqrt{N}}\sum_{\boldsymbol{k}}c_{\boldsymbol{k}}^{\alpha}e^{-i\boldsymbol{k}\cdot\boldsymbol{r}_{\hat{i}}}, \quad c_{\hat{i}}^{\alpha\dagger}=\frac{1}{\sqrt{N}}\sum_{\boldsymbol{k}}c_{\boldsymbol{k}}^{\alpha\dagger}e^{i\boldsymbol{k}\cdot\boldsymbol{r}_{\hat{i}}},
\end{equation}
in which $\alpha=\text{A, B}$. Under Fourier transformation, the Hamiltonian (\ref{hamiltonian4}) becomes
\begin{equation}
\label{hamiltonian5}
\mathcal{H}=\frac{1}{2}\sum_{\boldsymbol{k}}\psi_{\boldsymbol{k}}^{\dagger}\hat{\mathcal{H}}(\boldsymbol{k})\psi_{\boldsymbol{k}},
\end{equation}
in which we define the Nambu spinor $\psi_{\boldsymbol{k}}=\left(\begin{array}{cccc}
c_{\boldsymbol{k}}^{A}&c_{\boldsymbol{k}}^{B}&c_{-\boldsymbol{k}}^{A\dagger}&c_{-\boldsymbol{k}}^{B\dagger}
\end{array}\right)^{T}$, and the Hamiltonian matrix 
\begin{equation}
\label{hamiltonianmatrix}
\hat{\mathcal{H}}(\boldsymbol{k})=\left(\begin{array}{cccc}
-2J_{2}&B_{\boldsymbol{k}}&0&-A_{\boldsymbol{k}}^{*}\\B_{\boldsymbol{k}}^{*}&-2J_{1}&A_{-\boldsymbol{k}}^{*}&0\\0&A_{-\boldsymbol{k}}&2J_{2}&-B_{-\boldsymbol{k}}^{*}\\-A_{\boldsymbol{k}}&0&-B_{-\boldsymbol{k}}&2J_{1}
\end{array}\right),
\end{equation}
with the definition of complex variables
\begin{eqnarray}
\label{AandBvariables}
\begin{aligned}
&A_{\boldsymbol{k}}=-J_{1}\tau_{1}^{z}-iJ_{1}\tau_{2}^{z}+J_{2}\tau_{3}^{z}e^{i\boldsymbol{k}\cdot\boldsymbol{e}_{1}}-iJ_{2}\tau_{4}^{z}e^{i\boldsymbol{k}\cdot\boldsymbol{e}_{2}},\\ &B_{\boldsymbol{k}}=J_{1}\tau_{1}^{z}+iJ_{1}\tau_{2}^{z}+J_{2}\tau_{3}^{z}e^{-i\boldsymbol{k}\cdot\boldsymbol{e}_{1}}-iJ_{2}\tau_{4}^{z}e^{-i\boldsymbol{k}\cdot\boldsymbol{e}_{2}}.
\end{aligned}
\end{eqnarray}
Because we are only considering the states that have the lattice translational symmetry, the four $\tau^{z}$ fields in a unit cell are the only independent parameters of the distribution of the gauge fields. These and the strength of the coupling $J_{1}$ and $J_{2}$ are the input parameters in the Hamiltonian matrix (\ref{hamiltonianmatrix}). 

The Hamiltonian matrix $\hat{\mathcal{H}}(\boldsymbol{k})$ is Hermitian and it satisfies the following relation
\begin{equation}
\label{hrelation}
\hat{\mathcal{H}}(\boldsymbol{k})=-\tilde{\Lambda}\hat{\mathcal{H}}^{T}(-\boldsymbol{k})\tilde{\Lambda}, \text{ in which }  \tilde{\Lambda}=\left(\begin{array}{cc}
0&\hat{I}\\\hat{I}&0
\end{array}\right).
\end{equation}
This implies that if $E_{\boldsymbol{k}}$ is an eigenvalue of $\hat{\mathcal{H}}(\boldsymbol{k})$, $(-E_{-\boldsymbol{k}})$ is another eigenvalue of $\hat{\mathcal{H}}(\boldsymbol{k})$. Considering the definition of $\psi_{\boldsymbol{k}}$, this means that the Hamiltonian has particle-hole symmetry. The Hamiltonian (\ref{hamiltonian5}) belongs to the general class of BdG fermionic Hamiltonians. In order to study the diagonalization of the Hermitian matrix $\hat{\mathcal{H}}(\boldsymbol{k})$ in (\ref{hamiltonianmatrix}), we have  
\begin{equation}
\label{diagonalizingthehamiltonian}
\hat{\mathcal{H}}(\boldsymbol{k})=U^{\dagger}(\boldsymbol{k})\hat{\mathcal{D}}(\boldsymbol{k})U(\boldsymbol{k}),  \text{ with }  U^{\dagger}(\boldsymbol{k})U(\boldsymbol{k})=\hat{I}. 
\end{equation}
And the diagonal matrix takes the form $\hat{\mathcal{D}}(\boldsymbol{k})=\text{diag}(E_{1}(\boldsymbol{k}),E_{2}(\boldsymbol{k}),-E_{1}(-\boldsymbol{k}),-E_{2}(-\boldsymbol{k}))$.
Because of the particle-hole symmetry (\ref{hrelation}), we have, using the fact that $\hat{\mathcal{D}}(-\boldsymbol{k})=-\tilde{\Lambda} \hat{\mathcal{D}}(\boldsymbol{k})\tilde{\Lambda}$,
\begin{equation}
\label{Urelation}
U(\boldsymbol{k})=\tilde{\Lambda} U^{*}(-\boldsymbol{k})\tilde{\Lambda}.
\end{equation}
With these definitions, the Hamiltonian (\ref{hamiltonian5}) is then written as
\begin{equation}
\mathcal{H}=\frac{1}{2}\sum_{\boldsymbol{k}}\left(U(\boldsymbol{k})\psi_{\boldsymbol{k}}\right)^{\dagger}\mathcal{D}(\boldsymbol{k})\left(U(\boldsymbol{k})\psi_{\boldsymbol{k}}\right).
\end{equation}
We next define $\phi_{\boldsymbol{k}}=U(\boldsymbol{k})\psi_{\boldsymbol{k}}$, using (\ref{Urelation}) we can show that one can consistently write $\phi_{\boldsymbol{k}}=U(\boldsymbol{k})\psi_{\boldsymbol{k}}=\left(\begin{array}{cccc}
d_{\boldsymbol{k}}^{A}&d_{\boldsymbol{k}}^{B}&d_{-\boldsymbol{k}}^{A\dagger}&d_{-\boldsymbol{k}}^{B\dagger}
\end{array}\right)^{T}$, with $d_{\boldsymbol{k}}^{A}$ and $d_{\boldsymbol{k}}^{B}$ being fermionic operator satisfying $\{d_{\boldsymbol{k}}^{\alpha},d_{\boldsymbol{k}'}^{\beta}\}=0$ and $\{d_{\boldsymbol{k}}^{\alpha},d_{\boldsymbol{k}'}^{\beta\dagger}\}=\delta_{\boldsymbol{k}\boldsymbol{k}'}\delta_{\alpha\beta}$. Therefore the Hamiltonian (\ref{hamiltonian5}) is diagonalized as
\begin{equation}
\label{hamiltonianbcsdiagonal}
\mathcal{H}=\sum_{\boldsymbol{k}}E_{A}(\boldsymbol{k})d_{\boldsymbol{k}}^{A\dagger}d_{\boldsymbol{k}}^{A}+E_{B}(\boldsymbol{k})d_{\boldsymbol{k}}^{B\dagger}d_{\boldsymbol{k}}^{B}+\text{const}.
\end{equation}

Because the two gauge sectors are related under time reversal symmetry and thus have identical spectrum, we can only consider one sector, here we choose the sector $\tau_{1}^{z}\tau_{2}^{z}=1$. In this gauge sector, we can fix the values of the gauge fields to be $\tau_{1}^{z}=\tau_{2}^{z}=\tau_{3}^{z}=\tau_{4}^{z}=1$, due to gauge symmetry. For the next step, we fix our coordinate system for the lattice. Here we choose the lattice vectors to be (see Fig. \ref{figlattice} and Fig. \ref{figeffective})
\begin{equation}
\label{latticevectors}
\boldsymbol{e}_{1}=(-\frac{1}{2},\frac{\sqrt{3}}{2}), \qquad \boldsymbol{e}_{2}=(\frac{1}{2},\frac{\sqrt{3}}{2}).
\end{equation}
The reciprocal lattice vectors are thus $\boldsymbol{b}_{1}=4\pi(-\frac{1}{2},\frac{\sqrt{3}}{6}) \text{ and } \boldsymbol{b}_{2}=4\pi(\frac{1}{2},\frac{\sqrt{3}}{6})$. Furthermore we choose the Brillouin zone (BZ) as a rectangle $[-2\pi,2\pi]\times[0,\frac{2\sqrt{3}\pi}{3}]$ since it is simpler for numerical calculations. 

\begin{figure}
\includegraphics[width=0.4\textwidth]{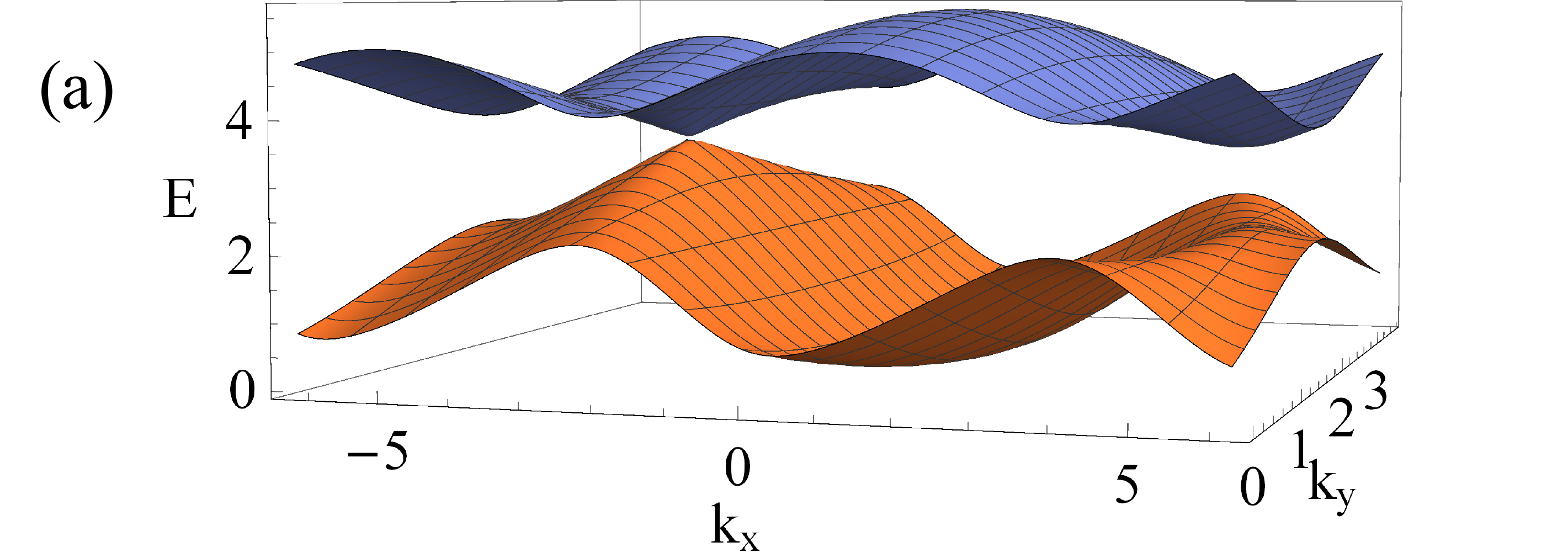}
\includegraphics[width=0.4\textwidth]{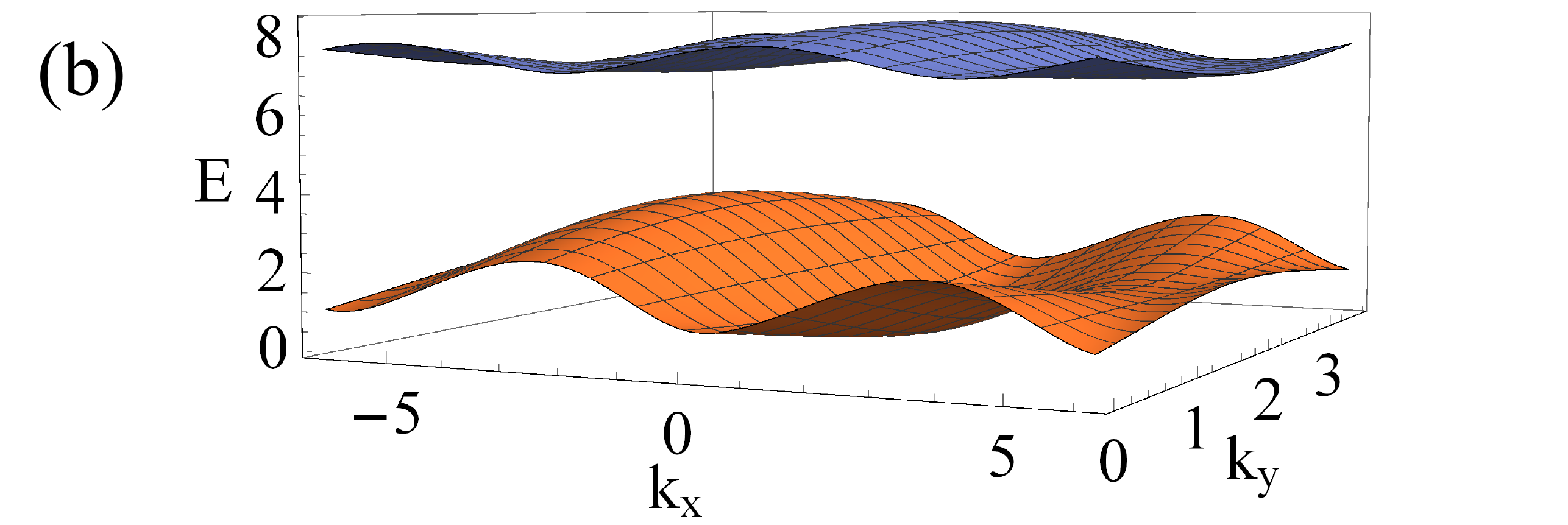}
\includegraphics[width=0.4\textwidth]{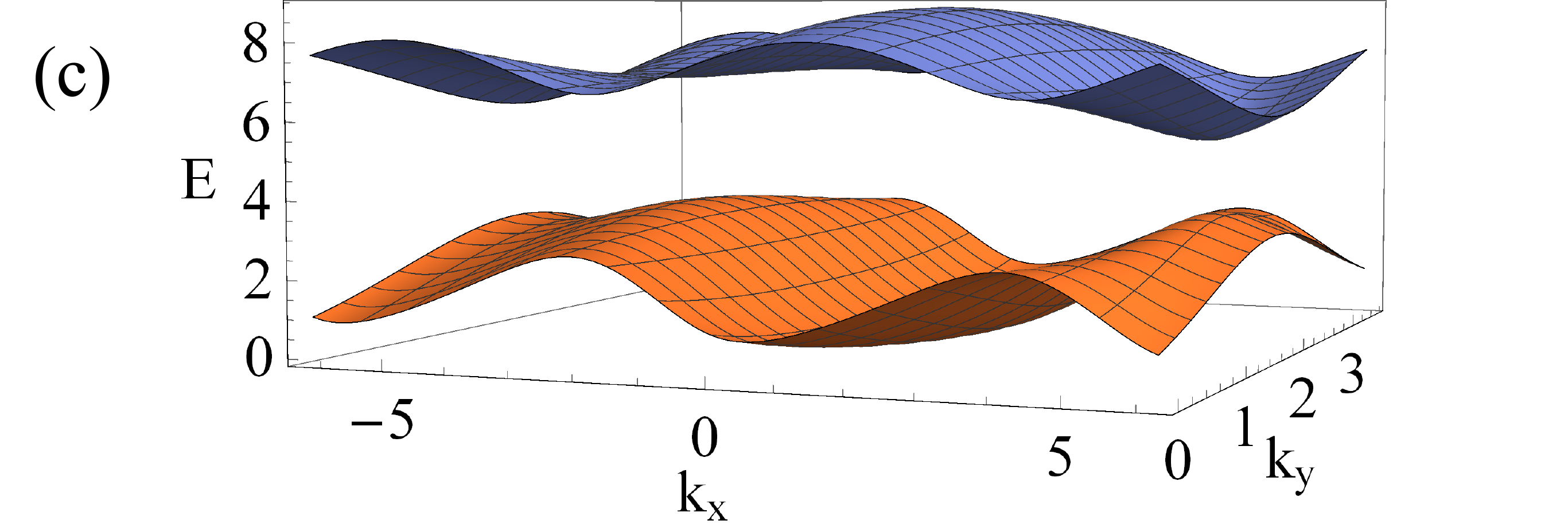}
\caption{The energy bands of the complex fermion for three different cases.(a) $J_{1}=1$, $J_{2}=1$. (b) $J_{1}=2$, $J_{2}=1$. (c) $J_{1}=1$, $J_{2}=2$. For all the three cases, the energy bands are computed under $\tau_{1}^{z}=\tau_{2}^{z}=\tau_{3}^{z}=\tau_{4}^{z}=1$ in every unit cell.}
\label{figenergyband}
\end{figure}

Among the four bands from the Hamiltonian matrix (\ref{hamiltonianmatrix}) only two are independent because of the particle-hole symmetry. After numerically diagonalizing the Hamiltonian matrix (\ref{hamiltonianmatrix}) with the input gauge fields $\tau_{1}^{z}=\tau_{2}^{z}=\tau_{3}^{z}=\tau_{4}^{z}=1$, we find that there are generally two positive energy bands and two negative energy bands. The ground state can thus be expressed as two empty positive bands. In the diagonalized Hamiltonian (\ref{hamiltonianbcsdiagonal}), suppose $E_{A}(\boldsymbol{k})>0$ and $E_{B}(\boldsymbol{k})>0$ (the other cases are related to this case with a constant unitary transformation on $\phi_{\boldsymbol{k}}$). According to the BCS theory of superconductivity \cite{simonsbook,BCS1957}, the ground state can be written as, 
\begin{equation}
|\Omega\rangle=\prod_{\boldsymbol{k}}d_{\boldsymbol{k}}^{A}d_{\boldsymbol{k}}^{B}|0\rangle,  
\end{equation}
with $|0\rangle$ being the ground state of the $c_{\boldsymbol{k}}$ fermions. Furthermore the two empty bands give the excitation spectrum of the model. In Fig. \ref{figenergyband} we plot the numerical results for the energy bands for $\tau_{1}^{z}=\tau_{2}^{z}=\tau_{3}^{z}=\tau_{4}^{z}=1$ in each unit cell with three different groups of coupling strength. 

From numerical calculation, it is found that there is always a $\boldsymbol{k}$-point at which the energy of the lower positive band is zero (see Fig. \ref{figenergyband}). For the lattice vectors we chose (\ref{latticevectors}), the location of this point is that $\boldsymbol{k}_{c}=(\frac{\pi}{3},\frac{\pi}{\sqrt{3}})$. This point has $\boldsymbol{k}_{c}\cdot\boldsymbol{e}_{1}=\frac{\pi}{3}$ and $\boldsymbol{k}_{c}\cdot\boldsymbol{e}_{2}=\frac{2\pi}{3}$. At this $\boldsymbol{k}$-point, it is straightforward to check that the determinate $\text{det}\hat{\mathcal{H}}(\boldsymbol{k}_{c})=0$ for any values of $J_{1}$ and $J_{2}$. This means that the excitation spectrum of the system is gapless. But there is always a gap between the lower positive band and the upper negative band.

\subsection{Generalization of the model with plaquette terms}

There are a number of conserved quantities in the spin model (\ref{hamiltonian1}), similar to the $W_{p}$ operators in the Kitaev model \cite{Kitaev2006,fu20181}. For the two basic plaquettes in the original wineglass lattice (see Fig. \ref{figlattice}), it is straightforward to define for the triangular plaquettes a conserved quantity $W_{3}=\sigma_{1}^{x}\sigma_{2}^{y}\sigma_{3}^{z}$; and for the nine-edge plaquettes a conserved quantity $W_{9}=\sigma_{1}^{x}\sigma_{2}^{x}\sigma_{3}^{x}\sigma_{4}^{z}\sigma_{5}^{z}\sigma_{6}^{z}\sigma_{7}^{y}\sigma_{8}^{y}\sigma_{9}^{y}$ (the numbering of the spins is not shown in the figures). After mapping to the Majorana fermions using the SO(3) Majorana representation, we can write these conserved plaquette quantities in terms of the $Z_{2}$ gauge fields. Specifically, we have 
\begin{equation}
W_{3}=\tau_{g}^{z}\tau_{h}^{z}.
\end{equation} 
And 
\begin{equation}
W_{9}=-\tau_{a}^{z}\tau_{b}^{z}\tau_{c}^{z}\tau_{d}^{z}\tau_{e}^{z}\tau_{f}^{z}.
\end{equation}
The labeling of the sites $a$ to $h$ is shown in Fig. \ref{figeffective}. Both $W_{3}$ and $W_{9}$ involve an odd number of spin operators and therefore they are both odd under time reversal. This agrees with the expressions with $Z_{2}$ gauge fields, in which an odd number of time-reversal odd gauge fields are involved in both cases. The terms $W_{3}$ and $W_{9}$ are $Z_{2}$ gauge invariant. Adding these terms to the Hamiltonian will break the time reversal symmetry explicitly, and therefore alters the energy spectrum of the states.

\section{Topological properties of the model and the Majorana zero mode} \label{secchernandzeromode}

\subsection{Chern number of the bands and gapless edge modes}

After the fixing of the $Z_{2}$ gauge fields, the Hamiltonian (\ref{hamiltonian5}) becomes a fermionic BdG Hamiltonian. This BdG Hamiltonian breaks time reversal symmetry. To see that, one simply notes the imaginary hopping coefficients in real space Hamiltonian (\ref{hamiltonian4}). According to the classification of topological phases of non-interacting fermions, this BdG Hamiltonian belongs to class $D$ \cite{chiu16,schnyder08,ludwig15,kitaev09}. It is characterized by an integer number, the Chern number \cite{tknn82,laughlin81,fradkinbook}. Physically the Chern number is defined as an integral of the Berry curvature over the BZ \cite{berry1984,xiao2010,aharonov1987,fradkinbook}, but here we use a more mathematical method to compute it \cite{Kitaev2006,chiu16,schnyder08,ludwig15}.

To find the Chern number that characterizes the topological properties of the spectrum, we define a matrix using the unitary matrix that diagonalizes the Hamiltonian matrix (\ref{diagonalizingthehamiltonian}),
\begin{equation}
Q=U^{\dagger}(\boldsymbol{k})\left(\begin{array}{cc}\hat{I}&0\\0&-\hat{I}
\end{array}\right)U(\boldsymbol{k}).
\end{equation}
This matrix is called the ``Q-matrix" \cite{chiu16,schnyder08,ludwig15}. It has the same eigenvectors as the Hamiltonian matrix itself (\ref{hamiltonianmatrix}). Furthermore, because the energy spectrum has two positive bands and two negative bands, the Q-matrix is directly related to the projector that projects the states to the occupied bands \cite{Kitaev2006}. The Q-matrix satisfies $Q^{\dagger}(\boldsymbol{k})=Q(\boldsymbol{k})$, $Q^{2}(\boldsymbol{k})=1$, and $\text{tr}Q(\boldsymbol{k})=0$. This means that the matrix $iQ(\boldsymbol{k})$ lives in both the Lie group U(4) and the Lie algebra $\mathbf{u}(4)$. For each unitary matrix $U(\boldsymbol{k})$, there is a corresponding $Q(\boldsymbol{k})$, but the Q-matrix is invariant under 
\begin{equation}
U(\boldsymbol{k})\rightarrow\left(\begin{array}{cc}
U_{1}&0\\0&U_{2}
\end{array}\right)U(\boldsymbol{k}), \qquad U_{1},U_{2}\in U(2).
\end{equation}
So the Q-matrix $Q(\boldsymbol{k})$ actually lives in the coset $U(4)/(U(2)\times U(2))$ \cite{chiu16,schnyder08,ludwig15}. 

The topological Chern number can be computed from the Q-matrix, it is given by \cite{chiu16,schnyder08,ludwig15,Kitaev2006}
\begin{eqnarray}
\label{chernnumber}
\begin{aligned}
\nu=&-\frac{i}{16\pi}\int \text{Tr}\left[Q dQ\wedge dQ\right]\\=&\frac{-i}{16\pi}\int \text{Tr}\left[Q\left(\frac{\partial Q}{\partial k_{x}}\frac{\partial Q}{\partial k_{y}}-\frac{\partial Q}{\partial k_{y}}\frac{\partial Q}{\partial k_{x}}\right)\right]dk_{x}dk_{y}.
\end{aligned}
\end{eqnarray}
For our model, the Chern number can be computed numerically after obtaining the Q-matrix for each $\boldsymbol{k}$-point in the BZ. We have numerically evaluated the Chern number for each combination of $\tau_{1}$, $\tau_{2}$, $\tau_{3}$ and $\tau_{4}$ and a range of coupling constants (both positive and negative) $J_{1}$ and $J_{2}$. We find that for all the cases the Chern number (\ref{chernnumber}) $\nu=\pm 1$ and furthermore, we obtain the following empirical formula for the Chern number
\begin{equation}
\label{chernnumberemperical}
\nu=-\text{sgn}(J_{1})\times (\tau_{1}^{z}\tau_{2}^{z}).
\end{equation} 
This formula is understandable according to the gauge structure of the model. The Chern number is equal for the gauge equivalent basis states, and it is odd under time reversal, therefore it only depends on the object $\tau_{1}^{z}\tau_{2}^{z}$ in a unit cell. 

The odd Chern number implies that the model has gapless chiral edge modes \cite{halperin82,fradkinbook}. For a wide range of the strength of the coupling constants, the Chern number only changes when $J_{1}$ changes sign. Moreover, for the spectrum of the fermion, there is always a gap between the lower positive band and the upper negative band. Therefore, there is no quantum phase transition in this model for different values of $J_{1}$ and $J_{2}$, in contrast with the case in the star lattice \cite{yao07,chung10}.

\subsection{Majorana zero mode associated with vortex excitations}

One key result for electronic systems with non-zero Chern number is that there is a Majorana zero mode associated with each vortex excitation \cite{volovik93,lee07,tewari07,roy10,teo2010,ivanov01}. For the Kitaev model itself, Kitaev gave two arguments for the existence of such Majorana mode with a flux excitation when the system is in the gapless phase under a magnetic field \cite{Kitaev2006}. The second argument is pretty general and goes beyond the translational invariant assumption. It states that there is an ``unpaired Majorana mode" associated with flux excitation when a generalized Chern number is $\pm 1$. Here we will follow his first argument and give a discussion about the existence of Majorana zero mode associated with flux excitation in our model.

This argument starts with taking another identical copy of the model with Majorana Hamiltonian (\ref{hamiltonian2}) and put it underneath the original model. One then pairs up the $\eta^{z}$ Majorana fermions on the same site of the two copies of the system. The result is an ordinary electron hopping model. Following this procedure, We first rewrite the Majorana Hamiltonian (\ref{hamiltonian2}) as, 
\begin{eqnarray}
\begin{aligned}
\label{hamiltonianmajorana}
\mathcal{H}=&\sum_{\hat{i}}\left[iJ_{1}\tau_{1}^{z}(\eta_{\hat{i}}^{3}\eta_{\hat{i}}^{1})+iJ_{1}\tau_{2}^{z}(\eta_{\hat{i}}^{3}\eta_{\hat{i}}^{2})+iJ_{1}(\eta_{\hat{i}}^{1}\eta_{\hat{i}}^{2})\right]+\\&\left[-iJ_{2}\tau_{3}^{z}(\eta_{\hat{i}}^{4}\eta_{\hat{i}+\boldsymbol{e}_{1}}^{2})-iJ_{2}\tau_{4}^{z}(\eta_{\hat{i}}^{4}\eta_{\hat{i}+\boldsymbol{e}_{2}}^{1})-iJ_{2}(\eta_{\hat{i}}^{3}\eta_{\hat{i}}^{4})\right],
\end{aligned}
\end{eqnarray} 
in which we keep the definition of the $Z_{2}$ gauge fields but use the Majorana fermions as the matter field. The numbering of the Majorana fermion is in Fig. \ref{figunit} (a) while the numbering of the $Z_{2}$ gauge fields is shown in Fig \ref{figunit} (b). All the Majorana fermion $\eta$ are $\eta^{z}$ in the original model, we omit the index $z$ for simplicity here. 

Now we introduce another layer of the same system, with the $Z_{2}$ gauge fields taking the same values as those in the original model (\ref{hamiltonianmajorana}) on corresponding sites. The matter fields in the new system are labeled by $\tilde{\eta}_{\hat{i}}^{\alpha}$ in unit cell $\hat{i}$ and position $\alpha=1,2,3,4$, corresponding to the Majorana field $\eta_{\hat{i}}^{\alpha}$ in the original layer. The two layers of the model do not have any coupling, we can simply write the total Hamiltonian as
\begin{equation}
\label{hamiltonianmajorana2}
2\mathcal{H}=\mathcal{H}(\tau,\eta_{\hat{i}}^{\alpha})+\mathcal{H}(\tau,\tilde{\eta}_{\hat{i}}^{\alpha}),
\end{equation}
in which the functional $\mathcal{H}(\tau,\eta_{\hat{i}}^{\alpha})$ is given by Eq. (\ref{hamiltonianmajorana}). Before we move on, we fix the gauge fields $\tau^{z}$ to be $+1$ throughout the lattice, which corresponds to picking up a basis state. It does not influence the final results because all the gauge equivalent basis states have the same properties and thus individual basis states behave like their corresponding physical states. 

We then pair up the corresponding Majorana fermions of the two identical models, $\eta_{\hat{i}}^{\alpha}$ and $\tilde{\eta}_{\hat{i}}^{\alpha}$ and define complex fermion $f_{\hat{i}}^{\alpha}$ such that $\eta_{\hat{i}}^{\alpha}=f_{\hat{i}}^{\alpha}+f_{\hat{i}}^{\alpha\dagger}$ and $\tilde{\eta}_{\hat{i}}^{\alpha}=i(f_{\hat{i}}^{\alpha}-f_{\hat{i}}^{\alpha\dagger})$. Under such definition, we have 
\begin{equation}
\eta_{i}^{\alpha}\eta_{j}^{\beta}+\tilde{\eta}_{i}^{\alpha}\tilde{\eta}_{j}^{\beta}=2(f_{i}^{\alpha\dagger}f_{j}^{\beta}+f_{i}^{\alpha}f_{j}^{\beta\dagger}).
\end{equation}
Using these we can write the total Hamiltonian of the two copies of the model (\ref{hamiltonianmajorana2}) as
\begin{eqnarray}
\label{hamiltoniancomplexf}
\begin{aligned}
&2\mathcal{H}=\\&2\sum_{\hat{i}}\bigg[iJ_{1}(f_{\hat{i}}^{3\dagger}f_{\hat{i}}^{1}-f_{\hat{i}}^{1\dagger}f_{\hat{i}}^{3})+iJ_{1}(f_{\hat{i}}^{3\dagger}f_{\hat{i}}^{2}-f_{\hat{i}}^{2\dagger}f_{\hat{i}}^{3})+\\&iJ_{1}(f_{\hat{i}}^{1\dagger}f_{\hat{i}}^{2}-f_{\hat{i}}^{2\dagger}f_{\hat{i}}^{1})\bigg]
+\bigg[-iJ_{2}(f_{\hat{i}}^{4\dagger}f_{\hat{i}+\boldsymbol{e}_{1}}^{2}-f_{\hat{i}+\boldsymbol{e}_{1}}^{2\dagger}f_{\hat{i}}^{4})\\&-iJ_{2}(f_{\hat{i}}^{4\dagger}f_{\hat{i}+\boldsymbol{e}_{2}}^{1}-f_{\hat{i}+\boldsymbol{e}_{2}}^{1\dagger}f_{\hat{i}}^{4})-iJ_{2}(f_{\hat{i}}^{3\dagger}f_{\hat{i}}^{4}-f_{\hat{i}}^{4\dagger}f_{\hat{i}}^{3})\bigg].
\end{aligned}
\end{eqnarray}
After Fourier transformation, $f_{\hat{i}}^{\alpha}=\frac{1}{\sqrt{N}}\sum_{\boldsymbol{k}}f_{\boldsymbol{k}}^{\alpha}e^{-i\boldsymbol{k}\cdot\boldsymbol{r}_{\hat{i}}}$, the complex fermion Hamiltonian can be written as $2\mathcal{H}=2\sum_{\boldsymbol{k}}\tilde{\mathbf{f}}^{\dagger}_{\boldsymbol{k}}\tilde{\mathcal{H}}(\boldsymbol{k})\tilde{\mathbf{f}}_{\boldsymbol{k}}$, in which the fermionic field $\tilde{\mathbf{f}}_{\boldsymbol{k}}=\left(\begin{array}{cccc}
f_{\boldsymbol{k}}^{1}&f_{\boldsymbol{k}}^{2}&f_{\boldsymbol{k}}^{3}&f_{\boldsymbol{k}}^{4}
\end{array}\right)^{T}$ and the matrix 
\begin{equation}
\tilde{\mathcal{H}}(\boldsymbol{k})=\left(\begin{array}{cccc}
0&iJ_{1}&-iJ_{1}&iJ_{2}e^{i\boldsymbol{k}\cdot\boldsymbol{e}_{2}}\\-iJ_{1}&0&-iJ_{1}&iJ_{2}e^{i\boldsymbol{k}\cdot\boldsymbol{e}_{1}}\\
iJ_{1}&iJ_{1}&0&-iJ_{2}\\-iJ_{2}e^{-i\boldsymbol{k}\cdot\boldsymbol{e}_{2}}&-iJ_{2}e^{-i\boldsymbol{k}\cdot\boldsymbol{e}_{1}}&iJ_{2}&0
\end{array}\right).
\end{equation}

One can perform the same analysis to the Hamiltonian matrix $\tilde{\mathcal{H}}(\boldsymbol{k})$, compute the Chern number associated with its spectrum using (\ref{chernnumber}). After numerical calculation, we find that the Chern number $\nu=-1$ for the spectrum of the complex fermion $f_{\hat{i}}$. This result can be easily understood considering the fermionic Hamiltonian (\ref{hamiltoniancomplexf}) as a real electronic Hamiltonian in a magnetic field. Specifically, based on the hopping coefficients on each bonds of the lattice, one can figure out the flux through each plaquette of the lattice; in this case, the magnetic flux are $\frac{\pi}{2}$ or $-\frac{\pi}{2}$ through every plaquette of the lattice (both the triangular plaquettes and the nine-edge plaquettes). The electronic system is in flux phases which is equivalent to a real magnetic field, therefore it breaks time reversal symmetry and the Chern number is non-zero.

Now we study the influence of a vortex excitation or flux excitation. Here we define a vortex excitation to be a plaquette around which the product of $Z_{2}$ gauge fields is $-1$. Note that in a finite system the flux excitations can only be introduced in pairs (one of the them can be on the boundary of the system) and the introduction of flux excitations breaks lattice translational symmetry. We suppose the system of the two identical layers of the original model has the shape of annulus, with a small hole in the center. In the ideal situation which we will adopt, the size of the hole is taken to be exactly one plaquette of the lattice. Having a flux excitation on the specific plaquette in both copies of the Majorana fermion model is equivalent to putting a magnetic flux $\pi$ through the hole in the complex fermion hopping model. As we know from the Integer Hall Effect, the Hall conductivity is given by \cite{tknn82,laughlin81,halperin82}
\begin{equation}
\label{TKNN}
\sigma_{xy}=\frac{e^{2}}{h}\nu,
\end{equation}
in which $\nu$ is the Chern number. In our case, $\sigma_{xy}=-\frac{e^{2}}{h}$. For the complex fermion system (\ref{hamiltoniancomplexf}), the Faraday effect of the induced magnetic flux will cause a net charge transferred to the edge of the hole or plaquette, as per Laughlin's argument \cite{laughlin81,thouless83}. In our case, the charge transferred is $\frac{e}{2}$. On the other hand, the $\pi$ flux is gauge equivalent to a $-\pi$ flux (for compact gauge groups as in our case), which is associated with a state with charge $-\frac{e}{2}$. These two charged states are equivalent to each other and thus have the same energy. Each individual state with extra charge $\frac{e}{2}$ or $-\frac{e}{2}$ breaks the compact gauge group symmetry of the complex fermion, the Goldstone mode between these two states is a fermion state with charge $e$ and zero energy, in other words, a fermion zero mode. Remembering that the complex fermion model is obtained from two identical copies of the Majorana fermion model with no coupling. Therefore there is a Majorana zero mode associated with each $\pi$ flux in the Majorana fermion model. The existence of isolated Majorana fermion is not a contradiction of physical laws because the $\pi$ flux can only be created in pairs in finite size systems.

In the argument above, when the $\frac{e}{2}$ charge is transported to the flux excitation, there is $-\frac{e}{2}$ charge transported to the edge of the system. This implies a connection between the Majorana zero mode and the edge states of the system with odd Chern number. As a matter of fact, as pointed out by Lee et al \cite{lee07}, the Majorana zero mode associated with a flux excitation can be understood as a Jackiw-Rebbi soliton \cite{jackiw76,su79,su80} associated with the one-dimensional chiral edge states. Finally, we note that the Majorana zero modes behave as Ising anyons when braiding among each other \cite{nayak2008,Kitaev2006,ivanov01}. We will not explore further about their properties as anyons.

\section{Conclusion and outlook} \label{secconclusion}

In this paper we have discussed the exact solution of a Kitaev type spin model on the triangle-star (wineglass) lattice with spontaneously broken time reversal symmetry. We show that the ground state of the model is a chiral spin liquid. The ground state of the model has Chern number $\nu=\pm 1$ indicating the existence of gapless chiral edge modes. However, unlike ordinary topological states of matter, the bulk excitation of the model itself is also gapless. Here, it is important to note that the gapless bulk in our model does not mean that the system is at a quantum critical point between two topological phases, as opposed to the cases such as the critical point of the $p_{x}+ip_{y}$ topological superconductor \cite{read00}. The key difference is that there is no band crossing over the entire BZ in our model, therefore there is no ambiguity in computing the Chern number. In other words, the topological phase in our model is stable although the bulk is gapless. From a broader perspective, the existence of a bulk gap is not necessary for the topological phases. For example, a group of models in which the gapless bulk and the gapless chiral edge modes coexist has been constructed by coupling a gapped topological matter with a gapless system\cite{baum20151,baum20152}. And it is shown that, for electronic systems of this kind, there are some experimental ways to distinguish the transport properties coming from the gapless bulk and the edge states \cite{baum20151,baum20152}. However, the fermonic degrees of freedom in our model are not electrons and they are not coupling to the electromagnetic fields as the ordinary electrons. To observe transport properties in our model, we have to rely on other techniques to generate the motion of the fermions. As proposed by Kitaev \cite{Kitaev2006}, one experimental technique that can be applied in our models is the thermal Hall effects \cite{kane97,katsura10}, in which the fermion motion is generated by thermal gradiant.

Due to the similarity between our model and the same type of model on the star lattice \cite{yao07,kells10,chung10}, it is expected that these two models show similar behavior. However, as opposed to the model on star lattice, there is no quantum phase transition when the relative strength of the spin coupling constants is varied. In the star lattice model, the topological phase with Chern number $\nu\neq 0$ only appears when the coupling on the edges of the triangles is large enough compared with the coupling on the other bonds connecting the triangles \cite{yao07}. The exact reason why there is no quantum phase transition on wineglass lattice as opposed to star lattice is an interesting subject for future study but here we can have some conjectures. Specifically the star lattice is made of triangle plaquettes and twelve-edge plaquettes, while the wineglass lattice in our model is made of triangle plaquettes and nine-edge plaquettes. The topological properties and time-reversal symmetry breaking result from the plaquettes with odd number of edges; therefore in the star lattice there may be a competation of phases resulting from coexistance of plaquettes with odd and even number of edges. On the contrary, there is no such competation in the wineglass lattice, hence no quantum phase transition. One has to consider other types of models of this kind on various kinds of lattices to confirm this conjecture. 

The model remains exactly solvable when the coupling constants on each bond is different from each other. Specifically, without breaking the lattice symmetry one can have different coupling constants in a unit cell, such as $J_{1,x}$, $J_{1,y}$ and $J_{1,z}$ on the triangle bonds and $J_{2,x}$, $J_{2,y}$ and $J_{2,z}$ on the star bonds. As the Kitaev model itself, there may be different phases associated with different relative values of these coupling constants, and also different kinds of excitations \cite{Kitaev2006}. Also, the bulk may remain gapless for a finite region in the phase diagram, just like the Kitaev model. Explorations in this direction is left for future study. Finally, one of the key problems is to find the model in real materials. The seminal work by Jackeli and Khaliullin \cite{Jackeli2009} has shed lights on the materials that host the Kitaev model. It would be interesting to find out if there is any materials that host the model in this paper.

\section*{Acknowledgements}

The author thanks the University of Minnesota for its support. This work is partly supported by NSF DMR-1511768 Grant.

\appendix

\section{Definition of $\tau^{x}$ fields in the Kitaev model} \label{appendixkitaev}

\begin{figure}
\includegraphics[width=0.4\textwidth]{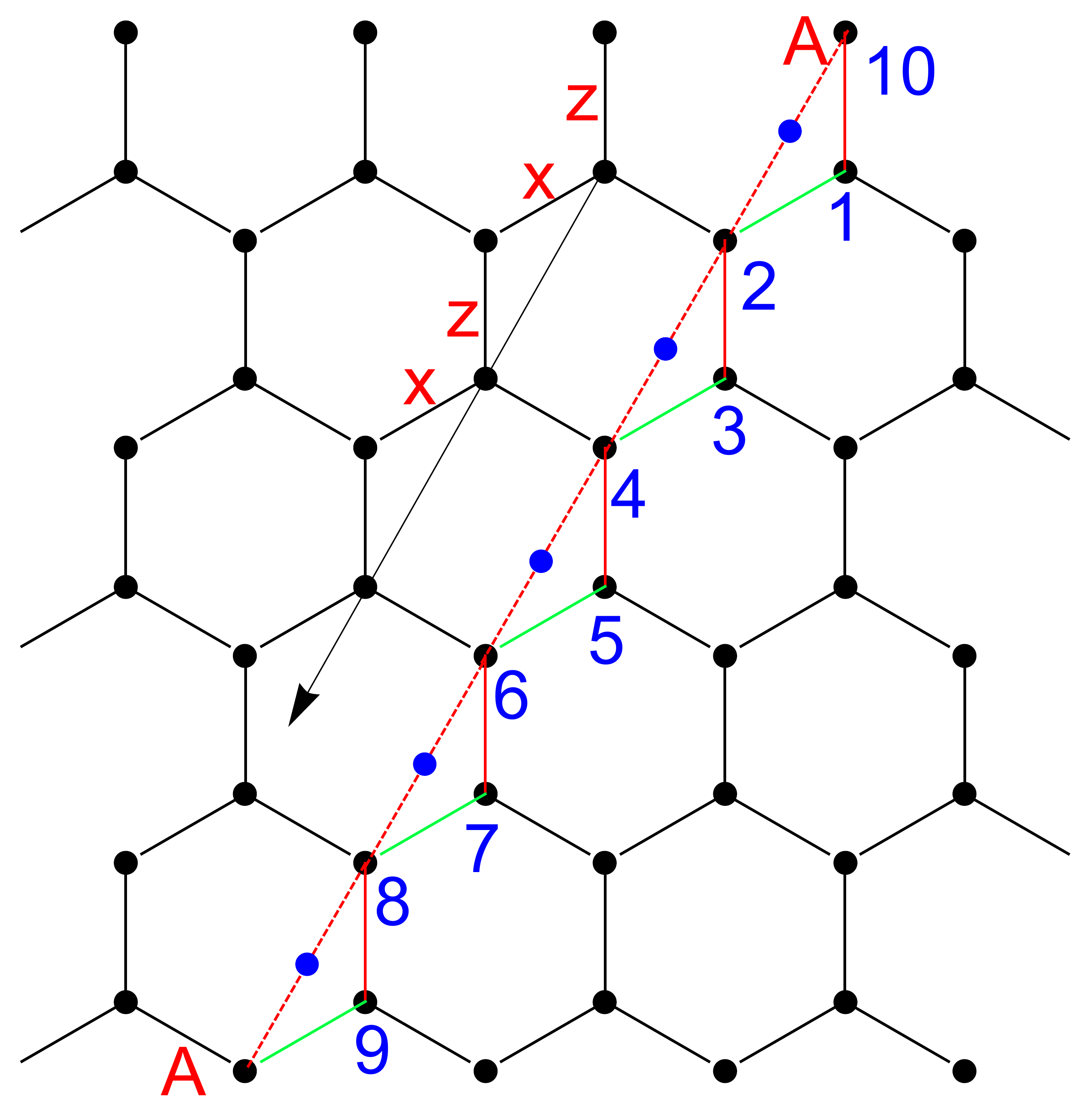}
\caption{The honeycomb lattice for the Kitaev model. To define the $\tau^{x}$ operators for $x$ type of bonds, one pick up a direction, given by the arrow. Here one of the $z$-bond-$x$-bond chain is shown with its sites numbered from 1 to 10. The lattice is finite with periodic boundary condition, with both sites A identified. }
\label{figkitaev}
\end{figure}

In this appendix, we briefly discuss how to define the $\tau^{x}$ gauge fields in terms of the Majorana fermions in the SO(3) Majorana representation solution of the Kitaev model \cite{fu20181}. This definition takes different form as the one used in Ref. \onlinecite{fu20181} and can be regarded as a more mathematical complement to that one. 

Following Ref. \onlinecite{fu20181}, in the solution of the Kitaev model using the SO(3) Majorana representation, the model is mapped into a lattice $Z_{2}$ gauge theory on the diamond lattice. In the diamond lattice, the $Z_{2}$ gauge fields on $x$ type of bonds are defined as 
\begin{equation}
\tau^{z}_{ij}\rightarrow i\eta_{i}^{y}\eta_{j}^{y}, \qquad \text{for $x$-bonds}.
\end{equation}
Although the positions of the $Z_{2}$ gauge fields are on the diamond effective lattice, here we label them by the original honeycomb lattice sites that it corresponds to. The corresponding $\tau^{x}$ fields can be defined in the following way. As shown in Fig. \ref{figkitaev}, we first pick up a direction on the lattice along $\boldsymbol{e}_{1}$ or $-\boldsymbol{e}_{1}$ (as denoted by the arrow in Fig. \ref{figkitaev}). For the specific $x$-bond $\langle ij\rangle$ on the honeycomb lattice (for example bond $\langle 56\rangle$), starting from the next $z$-bond along the chosen direction, we can obtain a zig-zag $z$-bond-$x$-bond chain (chain $6-7-8-9-\cdots$). For the finite size lattice, periodic boundary conditions make all the $z$-bond-$x$-bond chains loops; specifically in Fig. \ref{figkitaev}, there are ten sites on the loop. 

We then pick up a reference point on the chain (or loop for finite size systems), it has to be an end of a $x$-bond along the zig-zag chain (for example a reference point can be point 2 in Fig. \ref{figkitaev}). Therefore for each $x$-bond $\langle ij\rangle$, we obtain a specific $z-x-z-x-\cdots-z-x$ chain ending with the reference point, called chain $\langle ij\rangle$. The chain always includes odd number of lattice site (for bond $\langle 56\rangle$, the chain is $6-7-8-9-10-1-2$). Then we define the $\tau^{x}$ field on bond $\langle ij\rangle$ as
\begin{equation}
\tau^{x}_{ij}=\prod_{k\in \text{chain }\langle ij\rangle}\eta_{k}^{y}.
\end{equation}
It can be shown that the $\tau^{x}$ operators defined in this way satisfies $\{\tau_{ij}^{x},\tau_{ij}^{z}\}=0$ and $(\tau_{ij}^{x})^{2}=1$. It can also be shown that they are independent from other $Z_{2}$ gauge fields. 

Using the same procedure, one can define the $\tau^{x}$ operators in the wineglass lattice for the $Z_{2}$ Gauss law (\ref{constraintz2condition}) in Sec. \ref{secmodelandsolution}. This is how Eq. \ref{constraintz2condition} is obtained from Eq. \ref{constraint2}.

\bibliography{refJFU}

\end{document}